\documentclass[twoside,twocolumn]{article}
\usepackage[hmarginratio=1:1,margin={1.45cm,1.45cm}, top=32mm,columnsep=20pt]{geometry}
\usepackage[pdftex]{graphicx}
\usepackage{amsmath,amssymb,url,tikz,enumerate,siunitx,flushend}
\usepackage[normalem]{ulem}
\usepackage{booktabs}
\robustify\bfseries
\def\x{{\mathbf{x}}}

\def\xjhat{{\mathbf{\hat{x}}^j}}
\def\Y{{\mathbf{Y}}}

\def\Yjhat{{\mathbf{\hat{Y}}^{j}}}
\def\Mj{{\mathbf{M}^j}}

\def\V{\mathbf{V}}
\def\Vj{\mathbf{V}^{j}}
\def\Vjhat{\hat{\mathbf{V}}^{j}}
\def\Vin{\mathbf{V}_\text{in}}
\def\vin{\mathbf{v}_\text{in}}
\def\Vtr{\mathbf{V}_\text{tr}}
\def\vtr{\mathbf{v}_\text{tr}}
\def\rnnenc{RNN\textsubscript{enc}}
\def\rnndec{RNN\textsubscript{dec}}
\def\maskeroutput{\hat{\mathbf{V}}'^{j}_{\text{filt}}}
\def\denoiseroutput{\hat{\mathbf{V}}^{j}_{\text{filt}}}
\title{MaD TwinNet: \\Masker-Denoiser Architecture with Twin Networks \\for Monaural Sound Source Separation}

\author{%
Konstantinos Drossos$^{\dagger}$, Stylianos Ioannis Mimilakis$^{*}$, Dmitriy Serdyuk$^{\ddagger}$,\\ Gerald Schuller$^{*}$, Tuomas Virtanen$^{\dagger}$, Yoshua Bengio$^{\ddagger\S}$ \\
\\
$^{\dagger}$Tampere University of Technology, Tampere, Finland \\
$^*$Fraunhofer IDMT -- Technical University of Ilmenau, Ilmenau, Germany\\
$^{\ddagger}$MILA, Universit\'{e} de Montr\'{e}al, Montreal, Canada \\ 	 
$^{\S} $CIFAR Senior Fellow}
\date{}

\begin{document}
\maketitle
\begin{abstract}
Monaural singing voice separation task focuses on the prediction of the singing voice from a single channel music mixture signal. Current state of the art (SOTA) results in monaural singing voice separation are obtained with deep learning based methods. In this work we present a novel deep learning based method that learns long-term temporal patterns and structures of a musical piece. We build upon the recently proposed Masker-Denoiser (MaD) architecture and we enhance it with the Twin Networks, a technique to regularize a recurrent generative network using a backward running copy of the network. We evaluate our method using the Demixing Secret Dataset and we obtain an increment to signal-to-distortion ratio (SDR) of 0.37 dB and to signal-to-interference ratio (SIR) of 0.23 dB, compared to previous SOTA results. 
\end{abstract}

\section{Introduction}
Music source separation is an active research area in the context of audio signal processing and machine learning. The task is to estimate musical sources from observed musical mixture signals. One of the biggest challenges in music source separation is the estimation of singing voice from monaural (i.e. single channel) mixture signals~\cite{sisec17}. That is due to the high overlap that the sources exhibit in various signal representations~\cite{giannoulis11}.

Most approaches for source separation use time-frequency masking~\cite{mim17}. Modern state of the art method estimate the mask with neural networks of various architectures. Given the strong and long-term temporal patterns and structures of music (e.g. rhythm, beat/tempo, melody), architectures that can model long time dependencies, e.g. recurrent neural networks (RNNs), seem to be a great fit for the music source separation task. But, local structures are usually dominating the learning signal, because the RNN focuses on the most recent information~\cite{serdyuk:twinnet}. A known issue with the RNNs, that has been pointed out in many seminar works~\cite{serdyuk:twinnet}, e.g.~\cite{bengio:nn:1994,hochreiter:lsm:1997}. As a result, the long-term temporal patterns of music (e.g. tempo/beat, melody, and rhythm) might not be modeled correctly by an RNN, because the learning signal will be heavily influenced by the local structures~\cite{serdyuk:twinnet}. This means that the RNN will focus more on the local structures instead of the long-term temporal patterns~\cite{serdyuk:twinnet}.

The Twin Network (TwinNet)~\cite{serdyuk:twinnet} is an effective way to regularize generative RNNs when the generation is conditioned on some input (e.g. past content) and make the RNN also take into account the expected future content. This technique uses a second RNN which generates the same output in the backward direction and insures that the hidden states of the two networks are close. 

In this paper we are presenting a method that performs music source separation, using the TwinNet, and capable to model long-term structures of music. We evaluate the proposed method by focusing on singing voice separation (i.e. separating the singing voice from musical mixtures). This work builds upon the method for masking and denoising simultaneously~\cite{mim17}. The main contributions of this paper are:
\begin{enumerate}[i]
\item We present a method that improves the previous objective SOTA SDR and SIR by 0.37~dB and 0.23~dB, respectively. Our method is less computationally intensive comparing to the one presented in~\cite{mim17} which uses the recurrent inference (an iterative method which allows deep learning architectures to have stochastic depth \cite{stoch_depth});
\item We show that the TwinNet based regularization can be used for enhancing the results obtained by the MaD architecture.
\end{enumerate}
The rest of the paper is organized as follows. In Section~\ref{sec:overview} we give a brief overview of the related work for source separation task and approaches. The proposed method is thoroughly presented in Section~\ref{sec:proposedmethod}. The followed experimental procedure is described in Section~\ref{sec:experiments} and the obtained results are reported and discussed in Section~\ref{sec:results}. Section~\ref{sec:conclusions} concludes this work.

\section{Related work}
\label{sec:overview}
A common approach to estimate individual sources from monaural mixtures is to  to apply time-varying filters to the mixture signal~\cite{ps_masks}. The most straightforward way to derive and apply these filters is to treat audio signals as wide-sense stationary and compute a time-frequency representation of the signals via the short-time Fourier transformation (STFT). Then, the source estimate can be obtained by: 
\begin{equation}\label{eq:masking}
\Yjhat = \Y \odot \Mj \text{,}
\end{equation}
where $\Y$ and $\Yjhat \in \mathbb{C}^{M \times N}$ are the complex-valued STFT representations of the mixture signal vector $\x$ and the $j$-th source estimated signal $\xjhat$ respectively, with $M$ overlapping time frames and $N$ frequency sub-bands. $\Mj \in \mathbb{R}_{\geq 0}^{M \times N}$ is the $j$-th source-dependent filter, which we will refer to as \emph{mask}, and $\odot$ denotes the Hadamard product. The question that remains open is how to compute $\Mj$. 

In the case that all the $j$ sources are known \textit{a priori}, the source dependent masks are computed by employing ratios of the known sources' time-frequency representations~\cite{ps_masks,liutkus_alpha,voran17}. Selecting an appropriate method for mask computation when all the sources are known is outside the scope of this work. Interested readers are kindly referred to the following works~\cite{ps_masks, liutkus_alpha, voran17}. However, it is important to note that the mask computation is an \textit{open} optimization problem~\cite{fitz_masks} and in many cases assumptions about the source additivity~\cite{liutkus_alpha} and the phase dependencies~\cite{ps_masks, voran17} have to be made for many mask computations.

When the sources in an observed mixture are not known \textit{a priori}, supervised approaches relying on deep learning based optimization have yielded state-of-the-art results~\cite{sisec17}. Deep learning approaches for singing voice separation can be distinguished in three categories. The first category includes methods that train a deep neural network (DNN) to predict $\Mj$, conditioned on features computed using $\Y$~\cite{grais16} such as the magnitude spectrogram $\V = |\Y|$, where $|\cdot|$ denotes the matrix entry-wise absolute value operator. During training (when all sources are known for a given dataset), the pre-computation of the target $\Mj$ can rely on the ideal ratio mask (IRM, $\Mj_\text{IRM}\in[0,1]$), defined as
\begin{equation}
\mathbf{M}^{j}_\text{IRM}=\frac{\mathbf{V}^{j}}{\sum\limits_{j' \in J}\mathbf{V}^{j'}}\text{ where, }\label{eq:irm-mask}
\end{equation}
\noindent
$J$ is the total number of sources in a mixture. As can be seen, the form of the denominator implies the strong assumption that all the sources are additive. It must be noted here that the ideal amplitude mask (IAM, $\Mj_\text{IAM}\in\mathbb{R}_{\geq0}$), a usual alternative way of computing $\Mj$ is defined as
\begin{equation}
\mathbf{M}^{j}_\text{IAM}=\frac{\mathbf{V}^{j}}{\V}\text{, }
\end{equation}
\noindent
is considered to be not appropriate for deep learning approaches, due to the lack of upper limit of the mask~\cite{ps_masks}. As shown in~\cite{mim17_mlsp} and for tasks like singing voice separation, deep learning approaches that are trained to predict masks can be outperformed by methods that are not relying on pre-computed masks. The latter methods are discussed in the following paragraphs.

The second category follows the idea that was introduced in denoising autoencoders (DAEs)~\cite{vincent_den,bengio_den}. Specifically, DNNs are trained to recover the target source magnitude spectrogram $\Vj$ from a corrupted version of $\Vj$. The corrupted version of $\Vj$ is assumed to be the observed mixture magnitude spectrogram $\V$~\cite{uhl15, mim16}. For such methods, it was observed that the performance of the separation is highly dependent on post-processing steps that involved either the fusion of multiple trained DNNs \cite{uhl17} and/or post-processing steps such as masking of the mixture signal using the outcome of DNNs~\cite{mim17_mlsp, uhl15, uhl17, huang, takahashi17}.

Aiming to encapsulate the process of masking into deep learning optimization routines, the approaches of the third category introduced skip connections to the DNNs. The skip connections propagate the mixture signal $\V$ through two information paths. The first information path is the typical forward propagation of $\V$ through the layers of the DNNs and the second information path allows $\V$ to directly reach the output of the DNNs which is used to mask $\V$ using Eq.~(\ref{eq:masking}), yielding the final DNN estimate.

Specifically, the work~\cite{huang} employs deep RNNs and trained them to yield magnitude estimates for all the sources concurrently (i.e. $\Vjhat\;\forall\,j$). The magnitude estimates are then given to a deterministic function involving the computation of the ratio of the estimated magnitudes. The ratio outputs the mask that is applied to $\V$ and is encapsulated through the training procedure~\cite{huang}. That approach does not allow the deep RNNs to learn the masking process, but rather to output magnitude estimates that can be used to compute the mask. With the main ambition to also learn the masking process, the work of~\cite{mim16} proposed the usage of highway networks~\cite{hw15} that allow $\V$ to be masked directly by the output of a neural network layer. An extension to temporal sequences employing gated recurrent units (GRU)~\cite{bahdanau15} was presented in~\cite{mim17_mlsp} where the term skip-filtering connections was introduced. In~\cite{jannson17} a deep, ladder-structured, convolutional neural network (CNN) was presented. The output of the CNN was used to mask the input $\V$ to the CNN to provide magnitude estimates of the singing voice.

The limitations of the above methods are that highway networks of~\cite{mim16} do not compute any latent variables that can be used for denoising~\cite{vincent_den}. On the other hand, CNN architectures are prone to learn statistical irregularities of the data~\cite{jo17_cnns} making these two architectures not robust against small data perturbations~\cite{vincent_den, jo17_cnns}, while the GRU encoder-decoder of~\cite{mim17_mlsp} is not robust against interferences from other sources concurrently active in the mixture signal~\cite{mim17}. To tackle these problems, the \textit{masker-denoiser} (MaD) architecture was introduced in~\cite{mim17}. This architecture builds upon~\cite{mim17_mlsp} by incorporating a sparse transformation and a stochastic-depth optimization~\cite{stoch_depth} step that are used to generate the mask applied to $\V$ (i.e. the masker). As a final step a DAE with skip-filtering connections (i.e the denoiser) is responsible for eliminating remaining interferences from other music sources~\cite{mim17}.

\begin{figure*}[!ht]
\includegraphics[width=\textwidth]{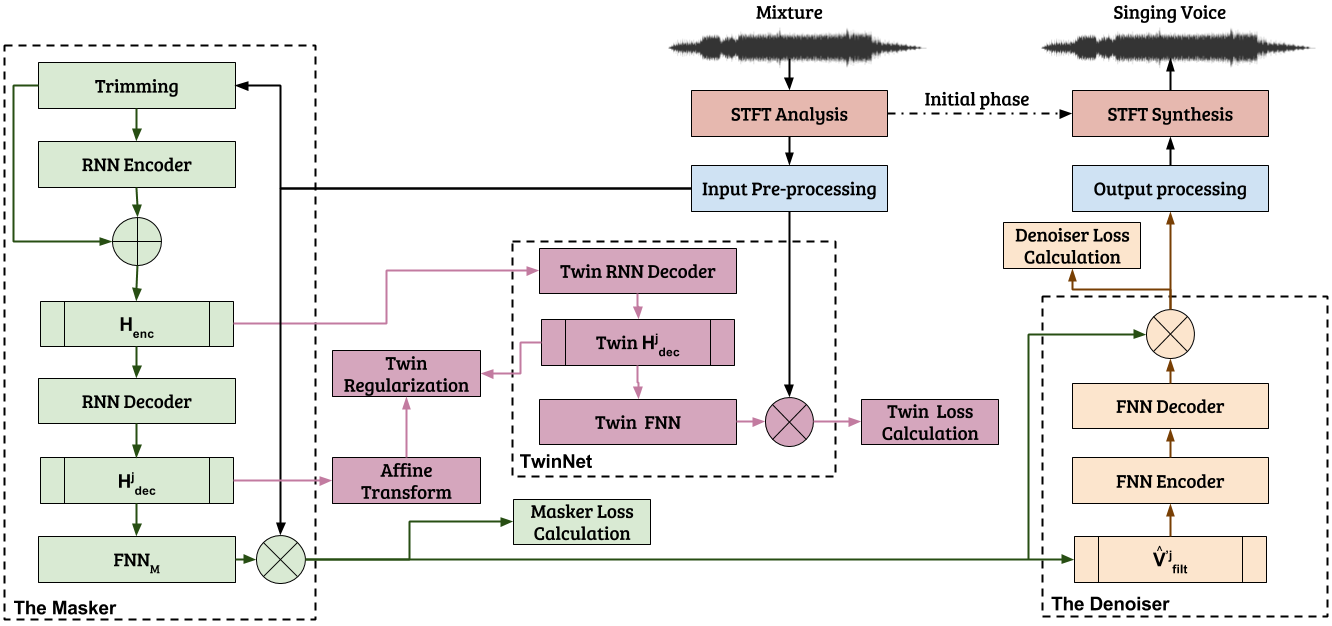}
\caption{Illustration of the proposed method. The parts in magenta color are used only during training.}
\label{fig:method_arch}
\end{figure*}

Although the approach in~\cite{mim17} provided state-of-the-art results in deep learning based monaural singing voice separation, the quality of the decoded mask is relying on stochastic optimization of a data-driven depth of the GRU decoder via recurrent inference~\cite{stoch_depth}. As a consequence and as the reported results in the experimental procedure of~\cite{mim17} suggest, the recurrent inference imposes a computationally cumbersome optimization process of the model given the available data for singing voice separation. 

To tackle that, we propose to replace the recurrent inference process by a recently developed Twin Network (TwinNet)~\cite{serdyuk:twinnet}. Similarly to the bidirectional RNN, the TwinNet employs a backward running network. The bidirectional RNNs are limited to be used for the representation learning when the TwinNet is applied for generative RNN. In addition to the original RNN, the TwinNet adds a second recurrent network that is a copy of the original with an exception that it aggregates the output in the backward direction. Second, the TwinNet adds a term to the loss function that depends on a trainable function of the backward running network. This loss function pushes together the hidden states of the forward net and the backward net for co-temporal timesteps. This cost ensures that the hidden state of the forward network encodes information stored in the state of the backward network. In other words, the TwinNet cost encourages the RNN to anticipate the future encoded in the backward running RNN, resulting in better modeling of both past and future context with the RNN. 

We use TwinNet under the hypothesis that the time-frequency mask for singing voice separation should remain the same regardless the direction (i.e. forward or backward in time) that one chooses to traverse a sequence of frames with time-frequency representation of the mixture signal. 

\section{Proposed method}
\label{sec:proposedmethod}
Our proposed method takes as an input the raw audio signal of a mixture of sources, and outputs the raw audio signal of the target source (i.e. the singing voice). The method uses a deep neural network architecture, which consists of two parts. The first part takes the mixture magnitude spectrogram of the input time-domain audio as an input, estimates the target source magnitude spectrogram from the mixture by predicting a time-frequency filter and applying it to the input, and outputs the estimated magnitude spectrogram of the target source. The second part takes the output of the first part as an input, predicts and applies a denoising mask, and outputs a filtered version of its input. Since the input to the first part is the mixture signal and the output is an estimate of the signal of the target source, we frame the time-frequency filter of the first part as a time-frequency mask. Similarly, since the input to the second part is a representation of the target source and its output is the same representation of the same source, we frame the time-frequency filter of the second part as a denoising filter. The first part of our method is denoted as {\em The Masker} and the second as the {\em The Denoiser}.

The Masker and the Denoiser are neural network architectures based on DAEs~\cite{vincent_den}. According to the initial paper of DAEs~\cite{vincent_den}, DAEs try to learn stochastically the manifold of the clean data. Source separation can be understood as a process that transforms the samples of the corrupted data manifold (i.e. the mixture) into samples that reside in the manifold of clean data (i.e. the target source)~\cite{den_ss}. This transformation for audio data can be understood using time-frequency masking~\cite{ps_masks}. We want to include in our optimization graph the generation of the mask and the denoising filter. Consequently, we setup our method in a way that the Masker will predict the mask that will be applied to the input magnitude spectrogram, and the denoiser will predict values of denoising filter which is applied to the output of the Masker. To do so, we employ the skip-filtering connections~\cite{mim17_mlsp}, which allow us to define the magnitude spectrogram of the target source as the target of the Masker and the Denoiser, but the output of the last neural network layer in the Masker and the Denoiser would be the mask and the denoising filter, respectively. 

We claim that the direction that one chooses to view the time-frequency representation of the mixture signal (i.e. forward or backward in time), does not affect the mask that is used in order to separate the singing voice from the mixture. Additionally, we hypothesize that a reverse traversing of the time-frequency representation of the mixture, will make the Masker to learn and anticipate the strong temporal patterns and structures of music. For these reasons, we use the recently proposed TwinNet for regularizing the Masker during training~\cite{serdyuk:twinnet}. Our method is illustrated in Figure~\ref{fig:method_arch} and thoroughly presented below. 

\subsection{Input pre-processing}
The input to our method is the vector of audio samples of the mixture signal $\mathbf{x}=[x_{1}, x_{2}, \ldots , x_{N}],\,\, x_{n} \in [-1,\,1]$, sampled at 44.1 kHz. We transform the input mixture signal into a time-frequency representation by using the STFT. For the STFT, we use overlapping frames of $N=2049$ samples ($\approx46$ milliseconds), segmented using the Hamming window function, and zero padded to $N'=4096$ samples. The hop size is set to 384 samples ($\approx9$ milliseconds). After the STFT, we retain only the first $N$ frequency sub-bands (i.e. up to the Nyquist frequency, including the DC term), resulting in the time-frequency representation of $\mathbf{x}$, $\mathbf{Y} \in \mathbb{C}^{M\times N}$, with $\V \equiv |\mathbf{Y}|$ to be the magnitude of $\mathbf{Y}$. From $\V$, we create overlapping subsequences of length $T$, that overlap by a factor of $L\times2$, in order to use context information from previous ($L$) and next (again $L$) frames. This results in having $B=\lceil{M/T}\rceil$ sequences of the form $\Vin \in \mathbb{R}^{T\times N}_{\geq0}$ for the whole input signal $\mathbf{x}$, where $\lceil\cdot\rceil$ is the ceiling function. Each $\Vin$ is used as an input to the next part of our method, which is the Masker. 

\subsection{The Masker}
The Masker consists of a frequency trimming process, a bi-directional recurrent neural network (Bi-RNN) encoder (\rnnenc), a forward RNN decoder (\rnndec), a sparsifying transform which is implemented with a rectified linear unit (ReLU) and a feed-forward neural network (FNN), and the skip-filtering connections. The input to the masker is the sequence $\Vin$ and its output is the predicted magnitude of the $j$-th target source, $\maskeroutput \in \mathbb{R}_{\geq0}^{T'\times N}$, with $T' = T-2L$.

$\Vin$ is trimmed to $\Vtr \in \mathbb{R}_{\geq0}^{T\times F}$, with $F=744$. This is done in order to reduce the dimensionality of the \rnnenc~and thus the number of training parameters. Consequently, information up to 8 kHz is retained. Since most of the relevant information for the singing voice is up to 8 kHz, the aforementioned reduction is considered to not have a great impact to the process of the Masker. 

$\Vtr$ is used as an input to the \rnnenc. The forward RNN of the \rnnenc~takes the sequence $\Vtr$ as an input, and the backward RNN the $\overleftarrow{\Vtr} = [{\vtr}_{T}, \ldots, {\vtr}_{t}, \ldots, {\vtr}_{1}]$. The hidden states of the forward and the backward RNNs of the \rnnenc~and at frame $t$, $\mathbf{h}_{t}$ and $\overleftarrow{\mathbf{h}_{t}}$ respectively, are concatenated and summed to the input (with residual connections) as
\begin{equation}
\label{eq:res_conn}
\mathbf{h}_{\text{enc}_{t}} = [(\mathbf{h}_{t} + {\vtr}_{t})^{\text{T}}, (\overleftarrow{\mathbf{h}_{t}} + \overleftarrow{\vtr}_{t})^{\text{T}}]^{\text{T}}\text{ ,}
\end{equation}
\noindent
leading to the output of the \rnnenc, $\mathbf{H}'_{\text{enc}} \in \mathbb{R}_{\geq-1}^{T\times2F}$,
\begin{equation}
\mathbf{H}'_{\text{enc}} = [\mathbf{h}_{\text{enc}_{1}}, \mathbf{h}_{\text{enc}_{2}}, \ldots, \mathbf{h}_{\text{enc}_{T}}]\text{ .}
\end{equation}
\noindent
Because we want the \rnndec~to focus only on the frames of $\mathbf{H}'_{\text{enc}}$ that are relevant to the sequence from which we want to extract the target source (i.e. the frames in the range $[1+L, T-L]$), we drop the first and the last $L$ $\mathbf{h}_{\text{enc}_{t}}$. This results to the output $\mathbf{H}_{\text{enc}} \in \mathbb{R}_{\geq-1}^{T'\times2F}$ of the \rnnenc,
\begin{equation}
\mathbf{H}_{\text{enc}} = [\mathbf{h}_{\text{enc}_{1+L}}, \mathbf{h}_{\text{enc}_{2+L}}, \ldots, \mathbf{h}_{\text{enc}_{T-L}}]\text{ .}
\end{equation}

$\mathbf{H}_{\text{enc}}$ is used as an input to the \rnndec, which outputs the hidden states $\mathbf{H}^{j}_{\text{dec}} \in [-1, 1]^{T',F}$. Consequently, the $\mathbf{H}^{j}_{\text{dec}}$ is used as an input to the sparsifying transform (the FNN with the ReLU) to obtain the time-frequency mask for the target source $j$, $\tilde{\mathbf{M}}^{j} \in \mathbb{R}_{\geq0}^{T'\times N}$, as
\begin{equation}
\label{eq:sparse-transform}
\tilde{\mathbf{M}}^{j} = g(\mathbf{H}^{j}_{\text{dec}}\mathbf{W}_{\text{FNN}_{\text{M}}} + \mathbf{b}_{\text{FNN}_{\text{M}}})\text{ ,}
\end{equation}
\noindent
where $\mathbf{W}_{\text{FNN}_{\text{M}}}$ and $\mathbf{b}_{\text{FNN}_{\text{M}}}$ are the weight matrix and bias vector of the FNN, respectively, and $g$ is the element wise ReLU activation. The output of the Masker is obtained by employing the skip-filtering connections as
\begin{equation}
\label{eq:skip-filtering-connection-masker}
\maskeroutput = \tilde{\mathbf{M}}^{j} \odot \Vin'\text{ ,}
\end{equation}
\noindent
where $\Vin'= [{\vin}_{1+L}, {\vin}_{2+L}, \ldots, {\vin}_{T-L}]$.

\subsection{TwinNet architecture and regularization}
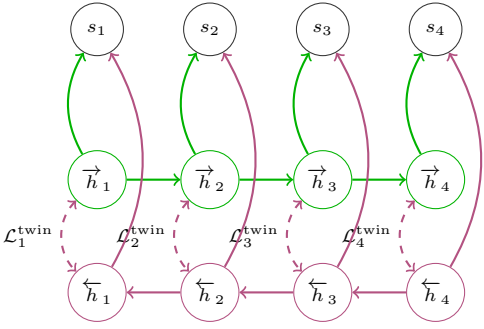
\begin{figure}
	\centering
    \begin{tikzpicture}[->,thick]
		\scriptsize
		\tikzstyle{main}=[circle, minimum size = 7mm, thin, draw =black!80, node distance = 12mm]
		\foreach \name in {1,...,4}
    		\node[main, fill = white!100] (y\name) at (\name*1.5,3.5) {$s_\name$};
		\foreach \name in {1,...,4}
    		\node[main, fill = white!100, draw=black!30!green] (hf\name) at (\name*1.5,1.5) {$\overrightarrow{h}_\name$};
		\foreach \name in {1,...,4}
    		\node[main, fill = white!100,draw=black!30!magenta] (hb\name) at (\name*1.5,0) {$\overleftarrow{h}_\name$};
		\foreach \h in {1,...,4}
    	{
       		\draw[<->,draw=black!30!magenta,dashed] (hf\h) to [bend right=45] node[midway,left] {$\mathcal{L}^{\text{twin}}_{\h}$} (hb\h)  {};
       		\path[draw=black!30!green] (hf\h) edge [bend left] (y\h);
    	}
		\foreach \current/\next in {1/2,2/3,3/4} 
    	{
       		\path[draw=black!30!green] (hf\current) edge (hf\next);
       		\path[draw=black!30!magenta] (hb\next) edge (hb\current);
    	}
		\foreach \h in {1,...,4}
    	{
       		\path (hb\h) edge [draw=black!30!magenta,bend right] (y\h);     
    	}
	\end{tikzpicture}
    \caption{Illustration of the hidden states regularization with the TwinNet. Forward RNN is depicted in green color. TwinNet (backward RNN) and TwinNet regularization are depicted in magenta color}
    \label{fig:twin}
    \bigskip
\end{figure}

The usage of a backward RNN as a regularizer of a forward RNN during training has been proposed in~\cite{serdyuk:twinnet,goyal2017z}. The target is to make the forward RNN capable to model better the long-term temporal structures and patters in the sequences that the RNN processes. To do so, the authors in~\cite{serdyuk:twinnet} use the hidden states of the backward RNN as a target for the hidden states of the forward RNN in a deterministic way, as can be seen in Figure~\ref{fig:twin}. 

More specifically, in the original proposal of the TwinNet~\cite{serdyuk:twinnet}, the authors consider a sequence of inputs $\mathbf{S} =[\mathbf{s}_{1},\ldots, \mathbf{s}_{T}]$, aiming at estimating the density $p(\mathbf{S})$. To do so, they target at maximizing the log-likelihood $\log p(\mathbf{S})$, using a forward RNN to process $\mathbf{S}$ and a non-linear transformation on top of the RNN, for predicting $p_{\text{f}}(\mathbf{s}_{t}|\mathbf{s}_{< t}) = \Psi_{\text{f}}(\overrightarrow{h}_{t})$. $\overrightarrow{h}_{t}$ is the output of the forward RNN and $\Psi_{\text{f}}(\cdot)$ is the non-linear transformation applied on top of the forward RNN (e.g. a softmax). 

To encourage the forward RNN to take into account upcoming (future) inputs, the authors in~\cite{serdyuk:twinnet} use a backward RNN to predict $p_{\text{b}}(\mathbf{s}_{t}|\mathbf{s}_{< t}) = \Psi_{\text{b}}(\overleftarrow{h}_{t})$, where $\overleftarrow{h}_{t}$ is the output of the backward RNN and $\Psi_{\text{b}}(\cdot)$ is a non-linear transformation applied on top of the backward RNN. Additionally, to regularize the learning process, they calculate the distance $\mathcal{L}^{\text{twin}}_{t} = ||f(\overrightarrow{h}_{t}) - \overleftarrow{h}_{t}||{}_2$, where $f$ is a learned affine transformation. All the above components all jointly optimized by maximizing
\begin{equation}\label{eq:twineqoriginal}
Q=\sum\limits_{t}\log p_{\text{f}}(\mathbf{s}_{t}|\mathbf{s}_{< t}) + \log p_{\text{b}}(\mathbf{s}_{t}|\mathbf{s}_{>t}) - \mathcal{L}^{\text{twin}}_{t}.
\end{equation}
\noindent
The $\mathcal{L}^{\text{twin}}_{t}$ at Eq.~(\ref{eq:twineqoriginal}), is the term that encourages the forward RNN to take into account the future inputs. 

During the evaluation/testing process, the backward RNN and the associated parts (i.e. the magenta parts in Figure~\ref{fig:twin}) are not used. Finally, in the optimization graph, all the preceding parts of the network from the backward RNN are not receiving a gradient signal from the objective of the backward RNN. This means that the input to the backward RNN is disconnected from the computation graph and is used only to optimize the backward RNN. 

In our work, we use the TwinNet only during training to regularize the $\mathbf{H}^{j}_{\text{dec}}$. We claim that the \rnndec~can greatly benefit from compensating for the future time frames, due to the strong temporal patterns and structures of music. We set up the TwinNet as a duplicate/twin of the \rnndec, the sparsifying transform, and the skip-connections, as can be seen in Figure~\ref{fig:method_arch}. The input to the TwinNet is the $\mathbf{H}_{\text{enc}}$ and its output is the $\mathbf{H}^{j}_{\text{twin}}$, calculated exactly as $\mathbf{H}^{j}_{\text{dec}}$. We apply
\begin{equation}
\label{eq:cost-twin}
\mathcal{L}^{\text{twin}} = \sum\limits_{t}||f(\mathbf{h}_{\text{dec}_{t}}) - \mathbf{h}_{\text{twin}_{t}}||{}_2
\end{equation}
\noindent
as the cost for the regularization of the \rnndec. The affine transform $f$ (obtained from a feed-forward network without any non-linearity applied to its output) is not used during evaluation/testing. In our work we transmit the gradient signal from the TwinNet back to the \rnnenc, in order to imbue the compensation for future values to the encoding part of the Masker as well. The output of the TwinNet that is used to optimize it is ${\Vjhat}_{\text{twin}}$ and is obtained exactly as $\maskeroutput$.

\subsection{The Denoiser}
We expect that the implemented masking process by the Masker will introduce artifacts to the magnitude spectrogram of the separated source. For that reason, we employ an extra learnable time-frequency filter applied to the $\maskeroutput$, in order to refine the latter and make it as close as possibly to the ${\Vj}'_{\text{in}}$. We perceive this process as denoising, hence we term the module that implements it the Denoiser. 

The Denoiser consists of two FNNs, the FNN\textsubscript{enc} and FNN\textsubscript{dec}, takes $\maskeroutput$ as an input, and outputs the $\denoiseroutput\in\mathbb{R}_{\geq0}^{T'\times N}$. The two FNNs of the Denoiser are set up in an DAE fashion and have shared weights through time. The FNN\textsubscript{enc} takes $\maskeroutput$ as the input and outputs $\mathbf{H}_{\text{FNN\textsubscript{enc}}}\in\mathbb{R}_{\geq0}^{T'\times N''}$, with $N'' = \lfloor N/2 \rfloor$ and $\lfloor\cdot\rfloor$ to be the floor function, as
\begin{align}
\mathbf{H}_{\text{FNN\textsubscript{enc}}} = g(\maskeroutput \mathbf{W}_{\text{FNN\textsubscript{enc}}} + \mathbf{b}_{\text{FNN\textsubscript{enc}}})\text{ ,}
\end{align}
\noindent
where $\mathbf{W}_{\text{FNN\textsubscript{enc}}}$ and $\mathbf{b}_{\text{FNN\textsubscript{enc}}}$ are the weights matrix and bias vector, respectively, of the FNN\textsubscript{enc}. The FNN\textsubscript{dec} accepts the $\mathbf{H}_{\text{FNN\textsubscript{enc}}}$ and outputs the $\mathbf{H}_{\text{FNN\textsubscript{dec}}}\in\mathbb{R}_{\geq0}^{T'\times N}$, as
\begin{align}
\mathbf{H}_{\text{FNN\textsubscript{dec}}} = g(\mathbf{H}_{\text{FNN\textsubscript{enc}}} \mathbf{W}_{\text{FNN\textsubscript{dec}}} + \mathbf{b}_{\text{FNN\textsubscript{dec}}})\text{ ,}
\end{align}
\noindent
where $\mathbf{W}_{\text{FNN\textsubscript{dec}}}$ and $\mathbf{b}_{\text{FNN\textsubscript{dec}}}$ are the weights matrix and bias vector, respectively, of the FNN\textsubscript{dec}. The final output of the Denoiser, $\denoiseroutput$, is obtained as
\begin{equation}
\label{eq:denoiser_output}
\denoiseroutput = \mathbf{H}_{\text{FNN\textsubscript{dec}}}\odot\maskeroutput\text{ .}
\end{equation}

\subsection{Output processing}
By iterating through all the overlapping subsequences of the analyzed input signal, the estimates from Eq. (\ref{eq:denoiser_output}) are aggregated together and reshaped to form the magnitude spectrogram of the $j$-th target source $\Vjhat \in \mathbb{R}_{\geq 0}^{M\times N}$. For the target source, we obtain the complex-valued representation by means of the Griffin-Lim algorithm (least squares error estimation from  modified STFT magnitude)~\cite{gla}, which uses the synthesis window and mixture's phase information. Inverse STFT is then applied to compute the time-domain samples of the target source $\mathbf{\hat{x}}^{j=1}$.

\subsection{Implementation and training details}
\label{subsec:training-details}
We jointly train all components of our method. We treat $\maskeroutput$, $\denoiseroutput$, and ${\Vj}_{\text{twin}}$ as matrices with unnormalized probabilities (i.e. the values are not summing up to one), allowing us to use the generalized Kullback-Leibler divergence as cost function, and the employed objective is
\begin{align}
\begin{split}
\mathcal{L} = &\mathcal{L}_{\text{D}} + \mathcal{L}_{\text{M}} + \mathcal{L}_{\text{TW}} + 0.5\mathcal{L}^{\text{twin}}\\ 
&+\lambda_{1}|\text{diag}\{\mathbf{W}_{\text{FNN}_{\text{M}}}\}|_{1}+\lambda_{2}||\mathbf{W}_{\text{FNN\textsubscript{dec}}}||^{2}_{2}\text{, where}
\end{split}\\
\mathcal{L}_{\text{D}} = &D_{\text{KL}}(\Vj \,\, || \,\, \denoiseroutput)\text{,} \\
\mathcal{L}_{\text{M}} = &D_{\text{KL}}(\Vj \,\, || \,\, \maskeroutput)\text{,}\\
\mathcal{L}_{\text{TW}} = &D_{\text{KL}}(\Vj \,\, || \,\, {\Vj}_{\text{twin}})\text{, and} 
\end{align}
\noindent
$\mathcal{L}_{\text{M}}$, $\mathcal{L}_{\text{D}}$, and $\mathcal{L}_{\text{TW}}$ are the objectives of the Masker, the Denoiser, and the TwinNet respectively, $D_{\text{KL}}$ is the generalized Kullback-Leibler divergence, $\lambda_{1}=\num{1e-2}$ and $\lambda_{2}=\num{1e-4}$ are regularization terms, $|\cdot|_{1}$ is the $\ell_{1}$ vector norm, and $||\cdot||_{2}^{2}$ is the $L_{2}$ matrix norm. $\text{diag}\{\mathbf{W}_{\text{FNN}_{\text{M}}}\}$ is the main diagonal of the weight matrix of the FNN of the Masker (i.e. the elements $w_{ij}$ of $\mathbf{W}_{\text{FNN}_{\text{M}}}$ with $i=j$). The values for $\lambda_{1}$ and $\lambda_{2}$ are after the vanilla version of the Masker-Denoiser architecture~\cite{mim17}. We employ the $\lambda_{1}|\text{diag}\{\mathbf{W}_{\text{FNN}_{\text{M}}}\}|_{1}$ in order to enforce the FNN of the Masker not to have energy in its main diagonal. We observed that high energy in the main diagonal of the $\mathbf{W}_{\text{FNN}_{\text{M}}}$ results to a source-dependent activity detector, rather than a source-dependent filter. We employ the $\lambda_{2}||\mathbf{W}_{\text{FNN\textsubscript{dec}}}||^{2}_{2}$ as a weight decay to avoid overfitting of the Denoiser and induce a sparsity factor. 

All RNNs are GRUs and are initialized using the orthogonal initialization technique from~\cite{saxe} and all other matrices using random samples from a normal distribution~\cite{glorot}. Biases are initialized with zeros. All parameters are jointly optimized using the Adam algorithm~\cite{adam}. The learning rate is set equal to $10^{-4}$, the batch size to $16$, and a gradient $L_2$ norm clipping equal to $0.5$ is applied. Out method is implemented using the PyTorch framework\footnote{\url {http://pytorch.org/}} and the code can be found online\footnote{\url{https://github.com/dr-costas/mad-twinnet}}.

\section{Experimental procedure}
\label{sec:experiments}
We assess the performance of our proposed method by focusing on the task of singing voice separation. We use the development subset of Demixing Secret Dataset\footnote{\url{http://www.sisec17.audiolabs-erlangen.de}} (DSD$100$) and the non-bleeding/non-instrumental stems of MedleydB~\cite{medleydb} for training our approach in a supervised fashion. Out total training set consists of 116 mixtures and their corresponding individual sources. The evaluation subset of DSD$100$ (50 mixtures and corresponding sources) is used for measuring the objective performance of our methods in terms of signal-to-distortion ratio (SDR) and signal-to-interference ratio (SIR), as proposed by the music source separation evaluation campaign (SiSeC)~\cite{sisec17}. Each multi-track contained in the available data is used to generate a monaural version of each of the four sources by averaging the two available channels. The target that we used for training (i.e. $\Vj$) is the outcome of the ideal ratio masking process~\cite{ps_masks}, scaled by a factor of $2$. The masking and the scaling processes are performed to avoid the inconsistencies in time delays and/or mixing gains between the mixture signal and the singing voice, which were apparent in the stems of the MedleydB dataset. Through our experiments it was observed that inconsistencies in the mixing gains yielded target source estimates that lacked amplitude, slightly decreasing the performance of the Masker. The length of the sequences is set to $T = 60$ (approximately equal to $0.5$ seconds), and the context information parameter to $L = 10$. The values for $T$ and $L$ are chosen after the initial proposal of the MaD architecture~\cite{mim17}.

We compared our proposed method with established SOTA approaches that solely deal with monaural singing voice separation. These approaches and their corresponding results are listed at the on-line results page of the signal separation evaluation campaign for music signals (SiSeC-MUS)\footnote{\url{https://sisec17.audiolabs-erlangen.de/#/results/1/4/2}}. The approach denoted as GRA3~\cite{grais16} is a DNN supervised approach to yield estimates for the ideal and/or IRM masks that are used to process the mixture magnitude spectrogram. The method denoted as CHA~\cite{cha17} is a CNN approach that yields estimates of all sources and then post-process them by using an IRM mask. STO2 is a DNN approach that operates on the common-fate signal representation~\cite{cmf}.

MM-RINF and MIM-NINF are MaD based methods, but {\emph{none of them uses the TwinNet regularization}. The MIM-RINF approach incorporates the recurrent inference stochastic optimization for the \rnndec, using a maximum number of $10$ iterations and a termination threshold equal to $\num{1e-3}$. The MIM-NINF approach does not incorporate the recurrent inference optimization procedure and it is the vanilla MaD architecture. Finally a supervised method based on robust principal component analysis (RPCA)~\cite{rpca17} for singing voice separation is also taken into consideration for the objective assessment, and this RPCA-based method is denoted as JEO2. The results from all the above mentioned approaches were obtained from the reported results of~\cite{sisec17} and~\cite{mim17}, following the same evaluation data and protocol proposed by SiSeC-MUS in \cite{sisec17}.

\section{Results}
\label{sec:results}
In Figures~\ref{fig:sdr-results} and~\ref{fig:sir-results} are the box plots for the obtained results, for the employed metrics (i.e. SDR and SIR), and compared with the previous SOTA results. In Table~\ref{tab:results} are the median values for the obtained results and compared with the same previous approaches. We use the median value because that is the one proposed and used by the SiSeC. An online demo of the separated audio sequences is available.\footnote{\url{http://arg.cs.tut.fi/demo/mad-twinnet/}}

\begin{figure}[!t]
\includegraphics[width=\columnwidth]{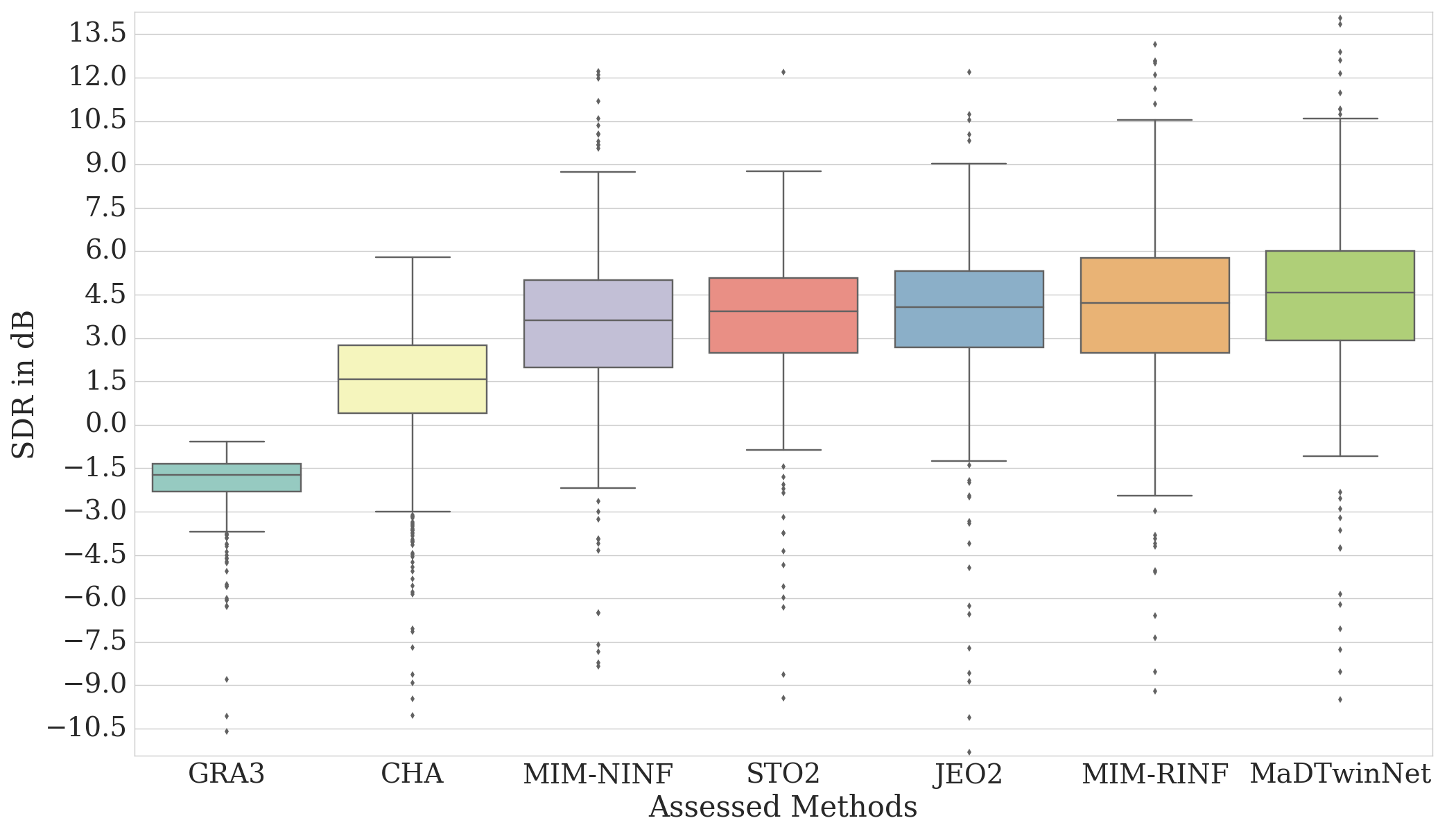}
\caption{Box-plots for the SDR obtained by MaDTwinNet and previous SOTA approaches.}
\label{fig:sdr-results}
\end{figure}

\begin{figure}[!t]
\includegraphics[width=\columnwidth]{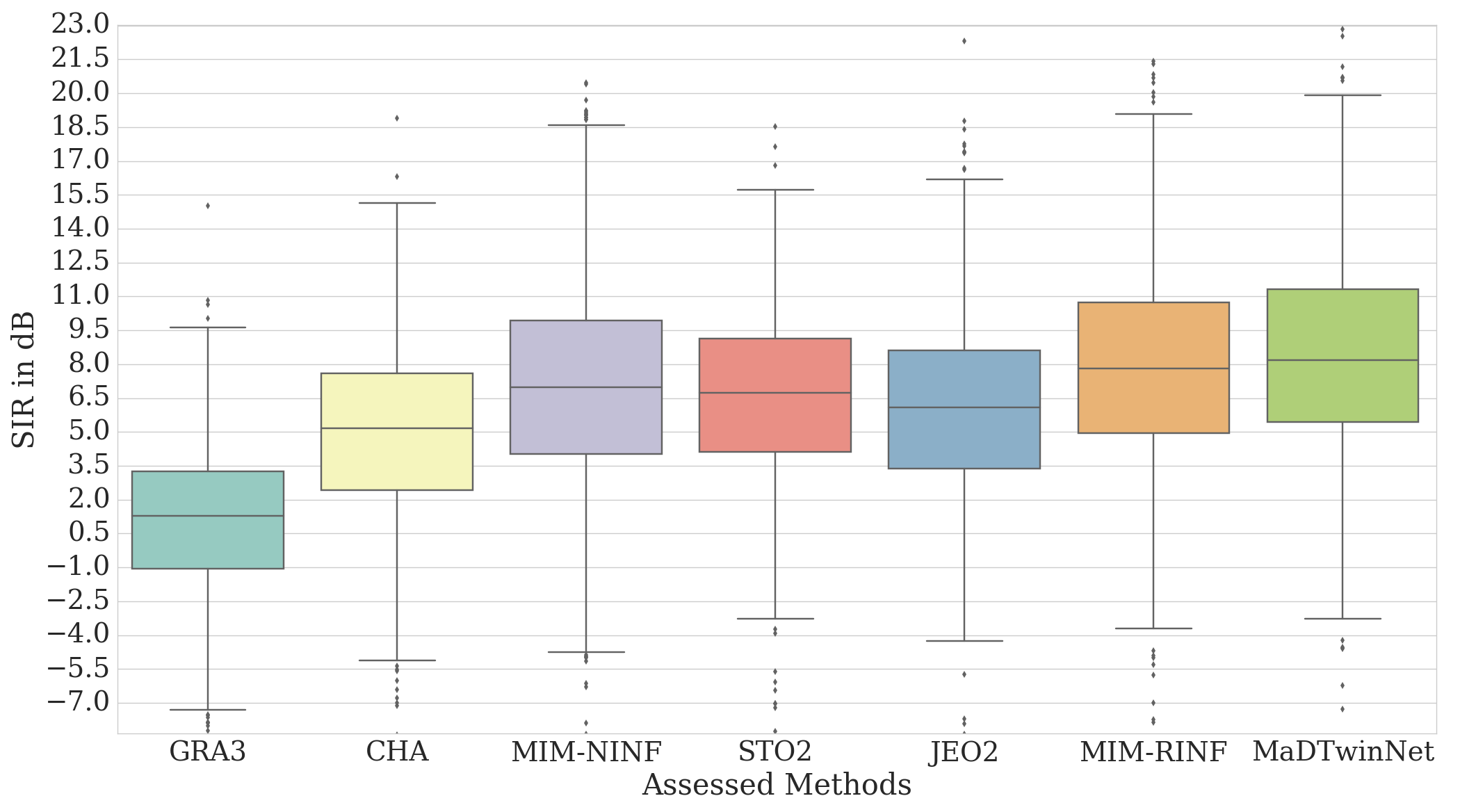}
\caption{Box-plots for the SIR obtained by MaDTwinNet and previous SOTA approaches.}
\label{fig:sir-results}
\end{figure}

\sisetup{detect-weight=true,detect-inline-weight=math}
\begin{table}[!t]
\centering
\caption{The median values for SDR and SIR of the proposed method and previous approaches.}
\label{tab:results}
\begin{tabular}{l
    S[
        table-format=1.2,
        table-space-text-pre={$\approx$}
	]
    S[
        table-format=1.2,
        table-space-text-pre={$\approx$}
	]}
\multicolumn{1}{c}{} & \multicolumn{2}{c}{\textbf{Metric}} \\
\multicolumn{1}{c}{\textbf{Approach}} & {\textbf{SDR}} & {\textbf{SIR}}\\
\midrule
GRA3 & -1.74 & 1.28\\
CHA & 1.58 & 5.17\\
MIM-NINF & 3.63 & 7.06\\
STO2 & 3.92 & 6.75\\
JEO2 & 4.07 & 6.09\\
MIM-RINF & 4.20 & 7.94\\
MaDTwinNet & \bfseries 4.57 & \bfseries 8.17\\
\end{tabular}
\end{table}

As can be seen in the Table~\ref{tab:results} and in the Figures~\ref{fig:sdr-results} and~\ref{fig:sir-results} for all the metrics, the MaD TwinNet method achieves higher scores than the previous approaches. Specifically, in Figure~\ref{fig:sdr-results} can be seen that the MaD TwinNet achieves better score from the previous better approach (i.e. MIM-RINF). At the same time, the MaD TwinNet has smaller range of values from the MIM-RINF approach, indicating more consistency at the expected results than the MIM-RINF approach. Since the basic difference from the MIM-* approaches is the TwinNet and the recurrent inference, the results for SDR indicate that the usage of the TwinNet leads to more robust methods. Additionally, MaD TwinNet surpasses in all aspects all other presented approaches in Figure~\ref{fig:sdr-results}. 

Almost the same trend can be observed at the results for the SIR, depicted at Figure~\ref{fig:sir-results}. Again, the MaDTwinNet surpasses all previous monaural approaches in the terms of the achieved SIR. The MaDTwinNet approach seems to yield higher SIR values, compared to the MIM-RINF approach. Compared with the results for the SDR, it can be seen clearly the the TwinNet regularization increase the performance of the MaD architecture. 

\section{Conclusions}
\label{sec:conclusions}
In this paper we proposed a method for music source separation, able to model both past and future context of a musical sound source. We augmented our previously proposed MaD architecture with the recently proposed TwinNet. The Masker (of the MaD architecture) outputs a first estimate of the magnitude spectrogram of the targeted source, and the Denoiser enhances this first estimate by removing artifacts introduced by the Masker. We used the TwinNet to regularize the Masker. We evaluated our proposed method using the free DSD dataset and focusing on the singing voice separation task. The results showed an increase to the previous obtained SOTA results on the same task. Specifically, we reached to an increase of 0.37~dB and 0.23~dB to SDR and SIR, respectively. The obtained results show that the TwinNet can enhance the performance of the MaD architecture.

As following work we propose the focus towards end-to-end methods, meaning that the neural network should receive as an input and produce as an output audio samples directly. This will make the time-frequency transformation to be included in the optimization graph and, probably, yield superior results. 

\section*{Acknowledgments}
Part of the computations leading to these results was performed on a TITAN-X GPU donated by NVIDIA to K. Drossos. K. Drossos and T. Virtanen wish to acknowledge CSC-IT Center for Science, Finland, for computational resources. D. Serdyuk would like to acknowledge the support of the following agencies for research funding and computing support: Samsung, NSERC, Calcul Qu\'{e}bec, Compute Canada, the Canada Research Chairs, and CIFAR. S.-I. Mimilakis is supported by the European Union's H2020 Framework Programme (H2020-MSCA-ITN-2014) under grant agreement no 642685 MacSeNet. The authors would like to thank P. Magron and G. Naithani (TUT, Finland) for their valuable comments and feedback during the writing process. 

\bibliographystyle{IEEEbib}
\bibliography{references}
\end{document}